# Effects of spatial nonlocality versus nonlocal causality for bound electrons in external fields

Ivan P. Christov

[1] Physics Department, Sofia University, 1164 Sofia, Bulgaria
[2] Institute of Electronics, Bulgarian Academy of Sciences, 1784 Sofia, Bulgaria

**Abstract:** Using numerically exact solution of the time-dependent Schrödinger equation together with time-dependent quantum Monte Carlo (TDQMC) calculations we compare the effects of spatial nonlocality versus nonlocal causality for the ground state and for real-time evolution of two entangled electrons in parabolic potential in one spatial dimension. It was found that the spatial entanglement quantified by the linear quantum entropy is predicted with good accuracy using the spatial nonlocality, parameterized naturally within the TDQMC approach. At the same time, the nonlocal causality predicted by the exact solution leads to only small oscillations in the quantum trajectories which belong to the idler electron as the driven electron is subjected to a strong high frequency electric field, without interaction between the electrons.

## 1. Introduction

There are generally two types of nonlocality which appear in non-relativistic quantum mechanics which are of different origin and have different implications on the phenomena under consideration. The one type manifests itself as a nonlocal causality which essentially means action-at-a-distance where influences can propagate even between non-interacting objects, irrespective of their spatial separation [1,2]. Basically, the nonlocal causality originates from the fact that whenever more than one quantum particles share the same wave function, it resides in a multi-dimensional configuration space which (although non-physical) binds the particles such that each of these may "feel" what happens to the rest. In quantum physics the nonlocal causality is closely related to the phenomenon of entanglement whereby the particles are correlated due to some kind interaction or due to symmetry conditions where they lose their individuality [3]. While in principle allowing faster-than-light influences the nonlocal causality is still compatible with special relativity owing to the statistical character of the measurement performed on multiple copies of identically prepared entangled systems [4]. For bipartite systems that type of quantum non-locality can be detected by e.g. testing violation of Bell's inequalities [5]. We do not consider here issues related to the instantaneous collapse of the wave function which accompany the measurement process or to those related to super-luminal propagation, whose description goes beyond the solution of the Schrödinger equation.

Another important type of nonlocality in quantum physics occurs in the form of convolution integrals which generally express the interaction of spatially extended objects, such as electron clouds, with other objects. A typical example in this respect is the mean-field potential in the familiar Hartree-Fock approximation where the electrostatic field at each point in space is an average over the whole charge distribution for a given electron (e.g. in [6]). Henceforth, we will call that type of nonlocality "spatial nonlocality" implying nonlocality in physical space-time. Although seemingly classical the spatial nonlocality considered here is intimately related to the quantum uncertainty and leads to entanglement in many-body quantum systems while other research attributes also the nonlocal causality to uncertainty [7].

A common feature of the two above types of nonlocality is that they usually assume some kind of preliminary interaction between the quantum particles to entangle their states in the transition from one-body to many-body state where, as a result, a nonfactorizable wave function is formed [8]. From computational viewpoint, however, it is known that the direct use of entangled many-body wave functions (which in general reside in multi-dimensional configuration space) is prohibitively expensive to compute due to both memory limitations and the exponential scaling of the workload. It is therefore of great importance to distinguish the above two mechanisms of quantum nonlocality and to compare their significance for doing efficient quantum calculations on a classical computer. In order to isolate and compare the observable effects of the two quantum nonlocalities in their pure form, we present here specific calculations where the spatial nonlocality and the nonlocal causality manifest themselves in an explicit and distinguishable manner. It is demonstrated that for the typical case of Coulomb interaction between electrons the influences that are due to nonlocal causality are at least two orders of magnitude weaker than those due to spatial nonlocality which may be important when evaluating the different mechanisms for entanglement in composite quantum systems such as molecules, clusters, and nano-structures.

## 2. Methods

The direct numerical solution of the Schrödinger equation alone is not sufficient to accomplish the tasks we discussed above since the calculated many-body wave function usually mixes the different sources of quantum nonlocality and entanglement for interacting particles. Therefore, for calculation of the effects that are due to only the spatial nonlocality (SN) we employ the recent time-dependent quantum Monte Carlo (TDQMC) method which accounts for the dynamic correlations that are due to e.g. the Coulomb repulsion between the electrons, but it basically ignores the nonlocal causality (NC) [9]-[11]. In fact, the lack of nonlocal causality is the price to be paid for achieving almost linear scaling with the system size in the TDQMC method against the exponential scaling of the original many-body Schrödinger equation.

Our calculations employ the often used model of one-dimensional two electron system in an external field, with a soft-core interaction between the electrons [12], where the Schrödinger equation reads (in atomic units):

$$i\frac{\partial}{\partial t}\Psi(x_1,x_2,t) = \left[H_0 + V_{ee}(x_1,x_2) + V_{ext}(x_1,x_2,t)\right]\Psi(x_1,x_2,t) \qquad (1)$$

where $H_0$ is the free Hamiltonian, which in our case is given by:

$$H_0 = -\frac{1}{2}\frac{\partial^2}{\partial x_1^2} - \frac{1}{2}\frac{\partial^2}{\partial x_2^2} + \frac{x_1^2}{2} + \frac{x_2^2}{2}, \qquad (2)$$

while $V_{ee}(x_1,x_2)$ is the repulsive soft-core interaction potential:

$$V_{ee}(x_1,x_2) = \frac{1}{\sqrt{1+(x_1-x_2)^2}} \qquad (3)$$

The additional potential $V_{ext}(x_1,x_2,t)$ in Equation 1 describes in our case the dipole interaction of the two electrons with external time-dependent electric fields $E_1(t)$ and $E_2(t)$ along the axes $x_1$ and $x_2$ of the two-dimensional configuration space, respectively:

$$V_{ext}(x_1,x_2,t) = -\left[E_1(t)x_1 + E_2(t)x_2\right] \qquad (4)$$

In order to evaluate the role of the spatial nonlocality in presence of external fields we solve numerically the set of quantum Monte-Carlo equations for two interacting electrons, which are represented as coupled Schrödinger-type of equations for a set of wave functions (guide waves) $\varphi_i^k(x_i,t)$ considered as random walkers in Hilbert space, together with a corresponding set of point walkers with trajectories $x_i^k(t)$ in physical space [11]:

$$i\frac{\partial}{\partial t}\varphi_i^k(x_i,t) = \left[-\frac{1}{2}\nabla_i^2 + V_{en}(x_i) + V_{eff}^k(x_i,t) + V_{ext}(x_i,t)\right]\varphi_i^k(x_i,t) \; ; \qquad i=1,2 \qquad (5)$$

where the core potential is $V_{en}(x_i) = x_i^2/2$ and $V_{eff}^k(x_i,t)$ is the effective interaction potential given by a Monte Carlo convolution of the true interaction potential $V_{ee}(x_i)$ and a kernel function $K\left[x_j, x_j^k(t), \sigma_j\right]$, which incorporates naturally the spatial nonlocality:

$$V_{eff}^k(x_i,t) = \sum_{j\neq i}^{N} \frac{1}{Z_j^k} \sum_{l}^{M} V_{ee}\left[x_i, x_j^l(t)\right] K\left[x_j^l(t), x_j^k(t), \sigma_j\right]; \qquad (6)$$

$$K\left[x_j, x_j^k(t), \sigma_j\right] = \exp\left(-\frac{\left|x_j - x_j^k(t)\right|^2}{2\sigma_j^2}\right), \qquad (7)$$

where i,j=1,2,…,$N$ is the number of electrons (two in our case), k=1,2,…,$M$ is the number of walkers for each electron, and $\sigma_j$ is the characteristic length of spatial nonlocality which has a clear physical meaning. Specifically, $\sigma_j$ quantifies the spatial range of each electron cloud which influences a given walker from another electron cloud such that the limit $\sigma_j \to \infty$ recovers the Hartree-Fock approximation while $\sigma_j \to 0$ corresponds to pairwise interaction between the walkers for the different electrons [10]-[11].

For real-time propagation the trajectories $x_i^k(t)$ are determined by the de Broglie-Bohm guiding equations for the velocity of each walker:

$$\frac{dx_i^k}{dt} = \operatorname{Im}\left[\frac{\nabla_i \varphi_i^k(x_i,t)}{\varphi_i^k(x_i,t)}\right]_{x_i = x_i^k(t)}, \qquad (8)$$

while these trajectories are determined by a drift-diffusion process during the ground-state preparation (imaginary-time $\tau$ propagation), according to:

$$dx_i^k(\tau) = v_i^{Dk} d\tau + \eta_i(\tau)\sqrt{d\tau}, \qquad (9)$$

where:

$$v_i^{Dk}(\tau) = \left.\frac{\nabla_i \varphi_i^k(x_i,\tau)}{\varphi_i^k(x_i,\tau)}\right|_{x_i=x_i^k(\tau)} \qquad (10)$$

is the drift velocity and $\eta_i(\tau)$ is a Markovian stochastic process. In this way, two ensembles of walkers in physical space and in Hilbert space are propagated self-consistently as particles $x_i^k(t)$ and guide waves $\varphi_i^k(x_i,t)$, respectively, where for imaginary-time propagation the walkers $x_i^k(\tau)$ sample the moduli square of the corresponding guide waves $|\varphi_i^k(x_i,\tau)|^2$, thus implementing the particle-wave dichotomy in physical space [11]. It is seen from Equations 5, 6 that for zero interaction potential between the elecctrons $V_{ee}(x_i)=0$ the effective potential is also zero and the equations for the different wave functions and trajectories are decoupled where these propagate independently from each other, thus ignoring the effects due to nonlocal causality.

For a purely variational calculation of the ground state we use here (no branching involved) the drift term in Eq.(9) is to be ignored while the system energy is calculated either using both trajectories and wave functions:

$$E_1 = \frac{1}{M}\sum_{k=1}^{M}\left[\sum_{i=1}^{N}\left[-\frac{1}{2}\frac{\nabla_i^2 \varphi_i^k(x_i^k)}{\varphi_i^k(x_i^k)} + V_{en}(x_i^k)\right] + \sum_{i>j}^{N} V_{ee}(x_i^k,x_j^k)\right]_{\substack{x_i^k=x_i^k(\tau)\\ x_j^k=x_j^k(\tau)}} ;i,j=1,2 \qquad (11)$$

or using wave functions only:

$$E_2 = \frac{1}{M}\sum_{k=1}^{M}\left[\sum_{i=1}^{N}\left[\frac{1}{2}\int|\nabla_i \varphi_i^k(x_i^k)|^2 dx_i^k + \int V_{en}(x_i^k)|\varphi_i^k(x_i^k)|^2 dx_i^k\right] + \sum_{i>j}^{N}\int V_{ee}(x_i^k,x_j^k)|\varphi_i^k(x_j^k)|^2 dx_j^k\right]; \quad i,j=1,2 \qquad (12)$$

In case of equal-spin electrons, additional terms should be added to Equation 6 and to Equations 11, 12 to account for the exchange interaction (see [13] for details). For real-time propagation the external potential in Equation 5 in dipole approximation reads $V_{ext}(x_i,t) = -E_i(t)x_i$.

### 3. Results

In order to isolate the effect of NC alone we first prepare the ground state of the two-electron system by solving the TDQMC Equations 5-10 for zero external fields using the split-step Fourier method over a spatial grid spanning 20a.u. The 1s ground state of two opposite-spin electrons is found by propagating two initial Gaussian distributions of width 1 a.u. in imaginary time until steady-state of the energy is established, for different values of the variational parameter $\sigma_j$, which physically determines the range of the spatial nonlocality as discussed above, (see [13] for details). The dependence of the system energy on $\sigma_1 = \sigma_2 = \sigma$ is depicted in Figure 1a where after carefully interpolating the original curve with a leat-squares polynomial fit (blue line) we find that the ground state energy approaches 1.7736a.u. for value of the variational parameter $\sigma = 0.82$ a.u.

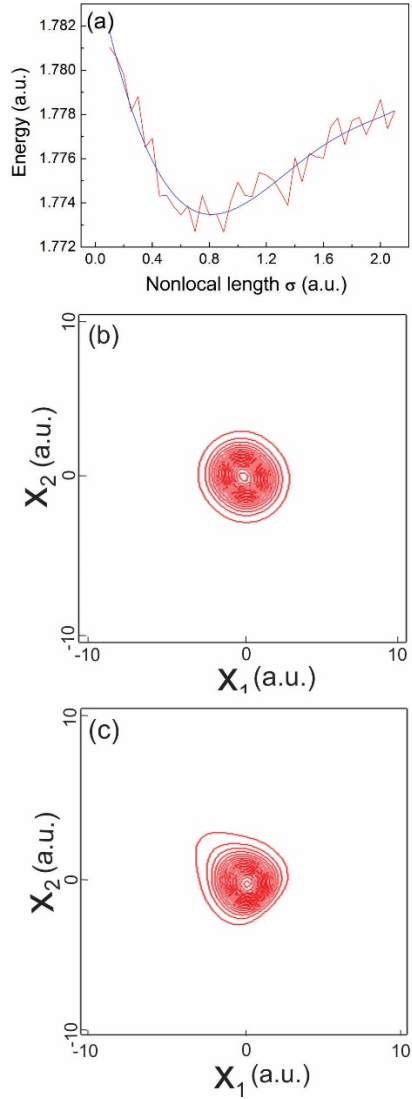

**Figure 1.** Energy of a two-electron system in 1s state as function of nonlocal length $\sigma$ -(**a**), ground-state wave function -(**b**), wave function after a few cycles of the sine electric field applied along the $x_2$ axis -(**c**).

Next, we find the numerically exact ground state by solving the two-body Schrödinger equation (Equation 1) in imaginary time using the same numerical method, for zero external potential, until stationary energy of 1.7735a.u. is established. In this way the "time-zero" two-body wave function $\Psi(x_1, x_2, t = 0)$, shown in Figure 1b, is found which is entangled due to the Coulomb repulsion between the electrons, which allows its further use to explore NC during real-time propagation. Since NC is expected to occur with no physical interaction between the electrons, at the second stage of our calculation we set the repulsion potential $V_{ee}$ to zero for $E_1 = E_2 = 0$ in Equation 4 which initially causes some collapse of the wave function towards the bottom of the two-dimensional parabolic potential of Equation 2. As that collapse is clearly independent on external fields we may simply subtract its effect from the result obtained in presence of the electric fields, which allows us to observe the pure motion of the wave function that is due to the electro-dipole interaction in Equation 1. During the real-time propagation we also monitor the dipole moments of the two electrons as induced by the external fields $E_1(t)$ and $E_2(t)$, together with a set of several random trajectories guided by the two-body wave function $\Psi(x_1, x_2, t > 0)$ according to [14]:

$$\frac{dx_2^k}{dt} = \text{Im}\left[\frac{\nabla_2 \Psi(x_1, x_2, t)}{\Psi(x_1, x_2, t)}\right]_{x_1 = x_1^k(t); x_2 = x_2^k(t)} \quad ; \quad k = 1, 2, 3\ldots \quad (13)$$

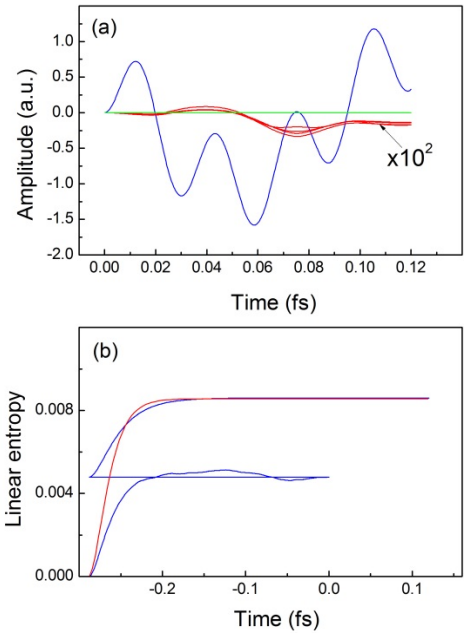

**Figure 2.** Trajectories and linear quantum entropies for non-interacting electrons: (**a**) – exact trajectories for the driven electron – blue line, for the idler electron – red lines, TDQMC result - green line; (**b**) – buildup and real-time dependence of the exact linear entropy – red line, and from the TDQMC calculation – blue line.

In order to observe the real-time effects of the nonlocal causality in our two-body system we set $V_{ee}(x_1, x_2) = 0$ for $t > 0$ and at the same time apply a sine electric field $E_1(t) = E_1^0 \sin(\omega_0 t + \varphi_0)$ with carrier frequency $\omega_0$=5a.u. and amplitude $E_1^0$=15a.u. to the first electron (called henceforth driven electron) while no electric field is applied to the other electron (idler electron, $E_2(t) = 0$ in Equation 4). It is seen from Figure 1c that when such strong electric field is applied to the driven electron, significant deformation in the shape of the two-body wave function $\Psi(x_1, x_2, t)$ takes place, which influences the idler electron even with no real interaction between the two electrons, which is exactly the effect of nonlocal causality in our case. Notice that the condition we use here that only one of the electrons is subjected to an external field is difficult to accomplish in practice for strongly overlapping 1s electrons as in a real atom. Nevertheless, that assumption seems justified if we want to clearly distinguish the effects of nonlocal causality from that of spatial nonlocality. Figure 2a depicts with blue line the dipole moment (essentially the average over all trajectories) for the driven electron while the red lines show a set of four separate trajectories which belong to the idler electron, as predicted by the exact solution of Equation 1 and Equation 13. These trajectories represent precisely the effect of the nonlocal causality in our case where we observe the influence between two non-interacting electrons through propagation of disturbances in the two-dimensional configuration space $(x_1, x_2)$ (see Figure 1c). It is important, however, that the net sum of all trajectories for the idler electron (essentially its induced dipole moment) approaches zero for all times which implies that no radiation by that electron due to the nonlocal interaction between the electrons may be expected. Similarly, the green line in Figure 2a corresponds to the result provided by the TDQMC calculation for the same parameters, where all trajectories are zero for all times due to the lack of nonlocal causality as discussed before.

As a measure for the entanglement caused by nonlocal interactions in the two-electron system, Figure 2b shown the linear quantum entropy [15] as predicted by the exact solution (red line) and that by the TDQMC method (blue line), where the latter uses density matrices built by the set of wave functions $\varphi_i^k(x_i, t)$ as described elsewhere [11]. It is seen from Figure 2b that for negative times (ground state preparation) the blue line consists of two stages where the first stage corresponds to solving the TDQMC Equations 5-7 in presence of random motion to the walkers $x_i^k(\tau)$ according to Equations 9 and 10, while the second stage repeats the imaginary-time propagation for fixed positions $x_i^k$ obtained after the first stage, in order to precisely adjust the wave functions $\varphi_i^k(x_i, t)$ to their steady-state positions. For the two-stage calculation it is more apropriate to estimate the ground-state energy using Equation 12 while for one-stage calculation both Equation 11 and Equation 12 can be used with close results. It is also seen from Figure 2b that the predictions of the TDQMC method for the linear entropy practically coincide with those of the numerically exact solution for both the ground state and the real-time propagation in presence of external fields (positive times in Figure 2b).

In order to account for the role of the spatial nonlocality for the real-time evolution we repeat the above calculations for non-vanishing interaction potential $V_{ee}(x_1, x_2)$. The result is seen in Figure 3a where the red and the green lines correspond to the dipole moment induced to the idler electron while the driven electron is a subject to the electric field, as in Figure 2a. A comparison between Figures 2a and 3a reveals that for non-interacting electrons (Figure 2a) the motion of the idler electron induced by the motion of the driven electron for positive times is almost two orders of magnitude smaller than in the case of interacting electrons (Figure 3a). The calculated quantum entropy for the case where both electrons experience the same external field is very close for the TDQMC and the exact solution as seen from the blue and the red curves in Figure 3b, where for clarity only the second stages of the ground state preparation are drawn. The light-blue and the green lines in Figure 3b show the linear entropies for values of the nonlocality parameter $\sigma$ equal to 0.7 a.u. and 1 a.u., respectively. Although these values are close to the optimal value of 0.82 a.u. of Figure 1a, the high sensitivity of the linear entropy on $\sigma$ seen in Figure 3b confirms our conclusion that the influence of the spatial nonlocality on the motion of the idler electron greatly exceeds that due the nonlocal causality.

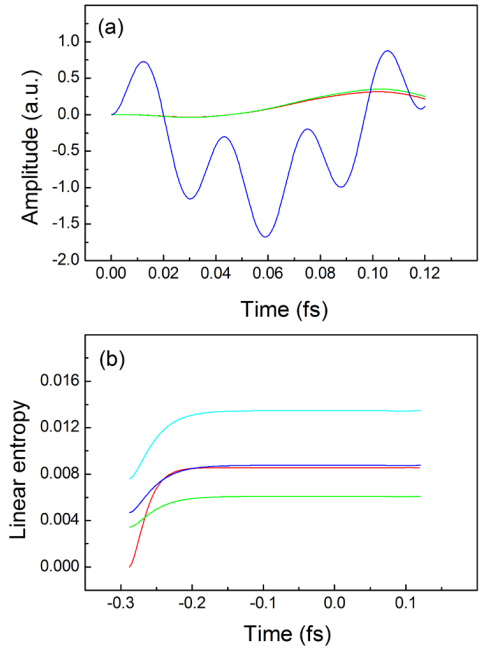

**Figure 3.** Trajectories and linear quantum entropies for interacting electrons: (**a**) – exact trajectories for the driven electron – blue line, for the idler electron – red line, TDQMC result - green line; (**b**) – buildup and real-time dependence of the exact linear entropy – red line, and the TDQMC result for: $\sigma = 0.82 a.u.$ – blue line, $\sigma = 0.70 a.u.$ - light-blue line, $\sigma = 1 a.u.$ - green line. In (b) both electrons are driven by the external field.

### 4. Conclusions

The aim of this paper was to compare the contribution of the nonlocal causality versus that of the spatial nonlocality to the motion of two electrons in 1s state. First, the entangled (due to the Coulomb repulsion) ground state is prepared and it is next propagated in real time while the interaction between the electrons is switched off and the one electron (driven electron) is subjected to powerful high frequency electric field. Under these conditions it was found from the numerically exact solution of the Schrödinger equation that the other electron (idler electron) experiences disturbances due to the nonlocal causality which is visualized through several random trajectories. In order to estimate the effect of the spatial nonlocality we conduct TDQMC calculations where the driven electron repels the idler electron through a weighted Coulomb potential which accounts for explicitly the spatial nonlocality. It was found that the amplitude of the idler electron's motion exceeds by two orders of magnitude its motion induced by the nonlocal causality alone. Both the spatial nonlocality and the nonlocal causality give rise to entanglement of the electron states which is quantified using linear quantum entropy. The amount of that entanglement is predicted correctly using spatial nonlocality alone, which supports our conclusion that the spatial nonlocality dominates over the nonlocal causality for both the ground state formation and the real-time evolution. Our results indicate that for doing practical calculations within reasonable accuracy one may solve Schrödinger equations in reduced dimensionality while ignoring the nonlocal causality, instead of pursuing more exact but expensive solutions.

**Funding:** This research is based upon work supported by the Air Force Office of Scientific Research under award number FA8655-22-1-7175, and by the Bulgarian Ministry of Education and Science as a part of National Roadmap for Research Infrastructure, grant number D01-401/18.12.2020 (ELI ERIC BG).